\documentclass[apj,iop]{emulateapj}

\usepackage{amsmath}
\usepackage{tabularx}
\usepackage{natbib}
\usepackage{float}
\usepackage[colorlinks=true,linkcolor=blue,citecolor=blue]{hyperref}%
\usepackage[all]{hypcap}
\usepackage{lineno}
\usepackage{footnote}

\shorttitle{Modeling 3D Polarization in Luhman 16 A and B}
\shortauthors{Mukherjee et al.}
\graphicspath{{./}{figures/}}

\begin{document}

\title{Modeling Polarization Signals from Cloudy Brown Dwarfs: Luhman 16 A and B in Three Dimensions}
\email{samukher@ucsc.edu}

\author{Sagnick Mukherjee$^{1}$, Jonathan J. Fortney$^{1}$, Rebecca Jensen-Clem$^{1}$, Xianyu Tan$^{2}$, Mark S. Marley$^{3}$, Natasha E. Batalha$^{4}$}
\affiliation{{$^1$}Department of Astronomy and Astrophysics, University of California, Santa Cruz, CA 95064, USA \\ 
{$^2$} Atmospheric, Oceanic and Planetary Physics, Department of Physics, University of Oxford, OX1 3PU, UK\\
{$^3$} Lunar and Planetary Laboratory, The University of Arizona, Tucson, AZ 85721, USA\\
{$^4$} NASA Ames Research Center, MS 245-3, Moffett Field, CA 94035, USA \\}










\begin{abstract}

The detection of disk-integrated polarization from Luhman 16A and B in H-band, and subsequent modeling, has been interpreted in the framework of zonal cloud bands on these bodies.  Recently, Tan and Showman (2021) investigated three-dimensional atmospheric circulation and cloud structures of brown dwarfs with general circulation models (GCMs), and their simulations yield complex cloud distributions showing some aspects of zonal jets, but also complex vortices that cannot be captured by a simple model.  Here we use these 3D GCMs specific to Luhman 16A and B, along with the three-dimensional Monte Carlo radiative transfer code ARTES, to calculate their polarization signals. We adopt the 3D temperature-pressure and cloud profiles from the GCMs as our input atmospheric structures.  Our polarization calculations at 1.6 $\mu$m agree well with the measured degree of linear polarization from both Luhman 16 A and B. Our calculations reproduce the measured polarization for both the objects with cloud particle sizes between 0.5-1 \,$\mu$m for Luhman 16 A and 5 \,$\mu$m for Luhman 16 B. We find that the degree of linear polarization can vary on hour-long timescales over the course of a rotation period.  We also show that models with azimuthally symmetric band-like cloud geometries, typically used for interpreting polarimetry observations of brown dwarfs, over-predict the polarization signal if the cloud patterns do not include complex vortices within these bands. This exploratory work shows that GCMs are promising for modeling and interpreting polarization signals of brown dwarfs.

\end{abstract}

\section{Introduction}\label{sec:intro}

Infrared spectroscopy of brown dwarfs has revealed a great deal of information about the thermal and chemical structure of their atmospheres \citep{kirkpatrick05}. These observations also have long suggested the presence of clouds in their atmospheres \citep{helling14,marley15}. However, cloud properties like their altitudes, vertical extent, and typical particle sizes are difficult to interpret from thermal spectroscopy or photometry. Moreover, time-series photometry and spectroscopy have often found these objects to be variable \citep{Buenzli15,apai21,milespaez15,eriksson19,milespaez17,artigau09,radigan12,girardin13,radigan14,vos19} suggesting non-uniform cloud coverage, which further complicates the interpretation of these observations.  

Additional observational and theoretical tools could potentially break important new ground in our understanding of substellar atmospheres.  One such avenue is polarization.  Thermal emission arising from substellar atmospheres may become partially linearly polarized due to scattering from the cloud/dust particles present in the atmosphere. Rayleigh scattering from gas molecules in the atmosphere can also give rise to significant polarization in optical wavelengths.  However, this gets suppressed in the near infrared due to the sharp decline of Rayleigh scattering with increasing wavelength \citep{marley11}. Therefore thermal polarized emission in near infrared wavelengths mainly arises from the Mie particle scattering of thermal photons by cloud particles that are present. 

However, the observable disk-integrated polarization from perfectly spherical and uniformly cloudy objects cancels out to zero even though the local polarized surface brightness of the object might be non-zero. A detectable non-zero disk-integrated polarization can still arise if this symmetry is broken either by rotationally-induced oblateness of the object or non-uniformity of the cloud cover \citep{senguptamarley10,marley11,sengupta01,dekok11}. For self-luminous gas giant planets, cousins of brown dwarfs, the presence of circumplanetary disks \citep{stolker17} or transiting exomoons \citep{senguptamarley16} can also break this symmetry for spherical objects and produce net disk-integrated polarized flux. 

Sensitivity of the disk-integrated polarization signal to the rotationally induced oblateness, inclination of the spin-axis of the object, gravity, effective temperature and cloud particle sizes has been theoretically studied in detail for uniformly cloudy exoplanets and brown dwarfs \citep{senguptamarley10,marley11,sengupta01,dekok11,sanghavi21,sanghavi19,sanghavi18}. Model calculations of disk-integrated polarization signals arising from non-uniform cloud coverage on exoplanets and brown dwarfs have been performed by \citet{dekok11} and \citet{stolker17}, yielding polarization signal predictions typically $\ge$0.1\% in the near-infrared wavelengths. At that level, near-infrared polarimetric and spectro-polarimetric observations of brown dwarfs can provide us with that  additional diagnostic window into the nature of clouds in these atmospheres.
 
A number of observations of polarized emission from brown dwarfs have been published as well over the past two decades \citep{menard02,Manjavacas17,zapatero05,zapatero11,goldman09,tata09,milespaez13,mmb20}. Recently, upper limits on the polarized thermal emission have been determined for a large sample (23) of exoplanets and brown dwarf companions \citep{jensenclem16,vanholstein17,jensenclem20,holstein21,mmb16} using the Gemini Planet Imager (GPI,\citet{Macintosh14}) and Spectro-Polarimetric High-contrast Exoplanet REsearch instrument (SPHERE, \citet{beuzit19}). \citet{holstein21} found that polarized thermal emission measurements from DH Tau B and GSC 6214-210 B hints towards the presence of circumsubstellar disks.
Time-domain photometric observations of some of the brown dwarf objects detected in polarized emission have additionally revealed significant variability in the thermal flux \citep{Buenzli15,apai21} and the polarized flux \citep{milespaez15}. This suggests not only the presence of clouds in these objects but might also hint towards non-uniformity in the cloud cover as well. However, no significant statistical correlation between the detection of polarized thermal emission and the presence of variability has been found yet. 
Nearly $\sim$40\% of brown dwarfs near the L--T transition are found to be strongly variable \citep{eriksson19} which suggests that these objects commonly have patchy cloud coverage, as had been suggested on theoretical grounds \citep{ackerman2001cloud,burgasser02}.

Our nearest binary brown dwarf pair Luhman 16 A and B \citep{luhman13} is especially interesting in this context, as Luhman 16 A has a spectral type of L7.5 and Luhman 16 B is a T0.5 dwarf \citep{kniazev2013,burgasser13}. As both of these objects are near the L--T transition,  there is a good chance that they have non-uniform cloud cover \citep{saumonmarley08}. Two dimensional Doppler mapping of Luhman 16 B by \citet{crossfield} found bright and dark regions across the globe which was indicative of cloud nonuniformity. \citet{Buenzli15} have shown that the thermal spectra of both the objects can be adequately fit with the superposition of two models with different cloud properties and coverage fractions. Multiple time domain photometric studies have also found both the components to be variable \citep{Buenzli15,apai21} further hinting towards patchy cloud coverage. 

\citet{mmb20} recently made the first detection of polarized thermal emission from each of the components of the binary system Luhman 16 A and B. The H-band degree of linear polarization for Luhman 16 A and Luhman 16 B was measured to be 0.031\% and 0.010\%, respectively with a precision of 0.004\%,  thus providing 7.5 and 2.5 sigma detections. The angle of the polarization vector was also measured for both the objects. Atmospheric modeling by \citet{mmb20} of the disk-integrated polarization signal showed that the measured polarization signal from Luhman 16 B can arise from a uniformly cloudy oblate object or from a non-uniformly cloudy configuration.  However, the polarization signal from Luhman 16 A could only be explained by non-uniform cloud coverage, for example multiple bands of relatively cloudy regions on the object \citep{mmb20}. This detection and the subsequent analysis emphasizes that non-uniformity in cloud cover should be considered while interpreting polarimetric (or even non-polarimetric) observations of brown dwarfs near the L--T transition. 

An essential tool to understand three-dimensional phenomena in substellar atmospheres are general circulation models (GCMs).  Such models couple fluid dynamics in three dimensions with radiative transfer, to understand energy transport throughout an atmosphere \citep{showman20}. A realization of the three-dimensional temperature and wind structure, when coupled with a treatment of cloud formation, yield physically-motivated predictions of substellar atmospheres. These GCMs provide the idealized basic structure of the atmosphere which can be post-processed with robust radiative transfer calculation codes like \texttt{ARTES}\citep{stolker17} enabling direct comparisons with polarimetric observations.

\citet{tanshowman21} and \citet{tanshowman20} have recently developed 3D circulation models for brown dwarfs that include a variety of cloud phenomena, like condensation of clouds and cloud radiative feedback, in order to simulate the global circulation and cloud patterns of rotating brown dwarfs. This gives us the opportunity to model the polarization signals emitted from these physically motivated, inherently 3D and non-uniform cloud distributions, to test if they can match the observations of Luhmam 16 A and B. In this work, for the first time we use 3D circulation models and post-process them with vector Monte Carlo radiative transfer code \texttt{ARTES} to interpret the polarimetric data, focusing on the observations of Luhman 16 A and Luhman 16 B from \citet{mmb20}. Despite the low statistical significance of the detection of polarization of Luhman 16 B, this object is of high interest to the astronomical community due to its proximity and significant variability, and we therefore consider it worthwhile to carry this modeling exercise on both A and B components. We specifically aim to explore and answer the following:
 
 \begin{enumerate}

      \item Can general circulation models with appropriate parameters for Luhman 16 A and Luhman 16 B be post-processed with a radiative transfer code like \texttt{ARTES} to match their observed polarization measurements?
      
      \item Do these GCMs also match the observed photometric variability in Luhman 16 A and B?
      
      \item Are the calculated polarization signal from circulation models of brown dwarfs sensitive to the spin-axis inclination relative to our line-of-sight of the objects?
      
      \item How sensitive is the disk-integrated polarization signal arising from GCM of brown dwarfs to the typical cloud particle sizes?
      
      \item Are cloud band models an adequate approximation when calculating polarization signals from objects which have non-uniform cloud covers including vortices of cloudy and clear regions?
 \end{enumerate}
 
We describe our modeling of the atmospheric circulation, cloud structures, and the polarized emission in \S \ref{sec:model}, results in \S \ref{sec:res}, discuss our findings in \S \ref{sec:disc}, and provide a summary and conclusion in \S \ref{sec:conc}.

%






\section{Modeling Sub-stellar Atmospheres}\label{sec:model}

\subsection{General Circulation Models of Brown Dwarfs}\label{sec:model_gcm}

We use the general circulation models (GCM) described and used in \citet{tanshowman20} and \citet{tanshowman21} to simulate the three-dimensional temperature and cloud structure for brown dwarfs Luhman 16 A and Luhman 16 B. The circulation models solve the fluid equations governing the horizontal momentum, mass continuity, hydrostatic balance and the energy transport of the substellar atmosphere in a rotating frame of reference. Two tracer equations are solved for the dynamics of the vapor and cloud components of the atmosphere. The radiative transfer for thermally emitted photons are computed using the gray atmosphere approximation in these models. The detailed physics and the governing equations of the GCM are fully described in \citet{tanshowman20}. We use the temperature structures and cloud distributions from these GCMs for calculating the disk-integrated polarized flux from Luhman 16 A and Luhman 16 B. Table \ref{table:planetparams} summarizes the physical input parameters used for simulating the circulation patterns for both the objects.

Of note, $N_c$ in Table \ref{table:planetparams} denotes the cloud particle number per dry air mass for each object.  The cloud material density used in the simulations corresponds to that of enstatite (MgSiO$_3$). $N_c$ controls the mean cloud particle sizes in each radial bin in our GCM according to Equation \ref{eq:particlesize} (see below) and the value of $N_c$ for each object shown in Table \ref{table:planetparams} is such that Luhman 16 A and B has typically 1 $\mu$m and 5 $\mu$m sized cloud particles, respectively, near the base of their cloud decks. We explore the effect of having different size cloud particles on the disk-integrated polarization signal from these objects in \S \ref{sec:res}.

The GCM models for Luhman 16 A use 60$\times$190$\times$384 grid cells corresponding to the $r,\theta,\phi$ variables, respectively, where $r$ is the radial dimension, $\theta$ is the polar dimension and $\phi$ is the azimuthal dimension. Models for Luhman 16 B use a 60$\times$254$\times$512 grid. The higher angular resolution used for Luhman 16 B is because it is a faster rotator, which requires following the dynamics on smaller length scales. The GCM model for both the objects evolve somewhat over time but here we use a time snapshot of the model after it reaches steady state. We now describe the modeled thermal and cloud structure from the GCM setups in \S \ref{sec:modeltp} and \S \ref{sec:modelcld}. 

\begin{table*}
\begin{center}

 \begin{tabular}{||c c c c c||} 
 
 \hline
 {\bf Parameter} & {\bf Model 16 A} & {\bf Model 16 B} & {\bf Measurement (16 A)} &  {\bf Measurement (16 B)} \\ [0.5ex] 
 \hline\hline
 Rotation Period &  7 Hours & 5 Hours  & 6.94 Hours \footnote{\label{apai}\citet{apai21}} & 5.28 Hours $\pm$ 22\% \textsuperscript{\ref{apai}}\footnote{\label{}Relative Period Range as reported in \citet{apai21}}\\ 
 \hline
 log(g) & 4.5   & 5  & 4.5\footnote{\label{buenzli15}\citet{Buenzli15}} & 4.5-5 \textsuperscript{\ref{buenzli15}} \\ 
 \hline
 Radius & 7$\times$10$^{7}$ m  & 6.3$\times$10$^{7}$ m & 0.95 R$_{\rm J}$\textsuperscript{\ref{buenzli15}} & 0.8-0.93 R$_{\rm J}$\textsuperscript{\ref{buenzli15}} \\
 \hline
 $N_c$ & 1$\times$10$^{9}$ kg$^{-1}$  & 1$\times$10$^{8}$ kg$^{-1}$ & -- & --\\
 \hline
 Cloud Material Density & 3.19 g/cm$^{3}$  & 3.19 g/cm$^{3}$ & -- & -- \\
 \hline
 Temperature (100 Bars) & 4300 K & 3500 K & -- & --\\
 \hline
\end{tabular}
\end{center}
\caption{The brown dwarf parameters and their values used to model the three-dimensional circulation patterns for Luhman 16 A and Luhman 16 B are tabulated. The measured values of these parameters are also shown except for those parameters which are not yet measured but has been assumed in our modeling. }
\label{table:planetparams}
\end{table*}

\subsubsection{The 3D Temperature-Pressure Structure}\label{sec:modeltp}
\begin{figure*}
  \centering
  \includegraphics[width=1\textwidth]{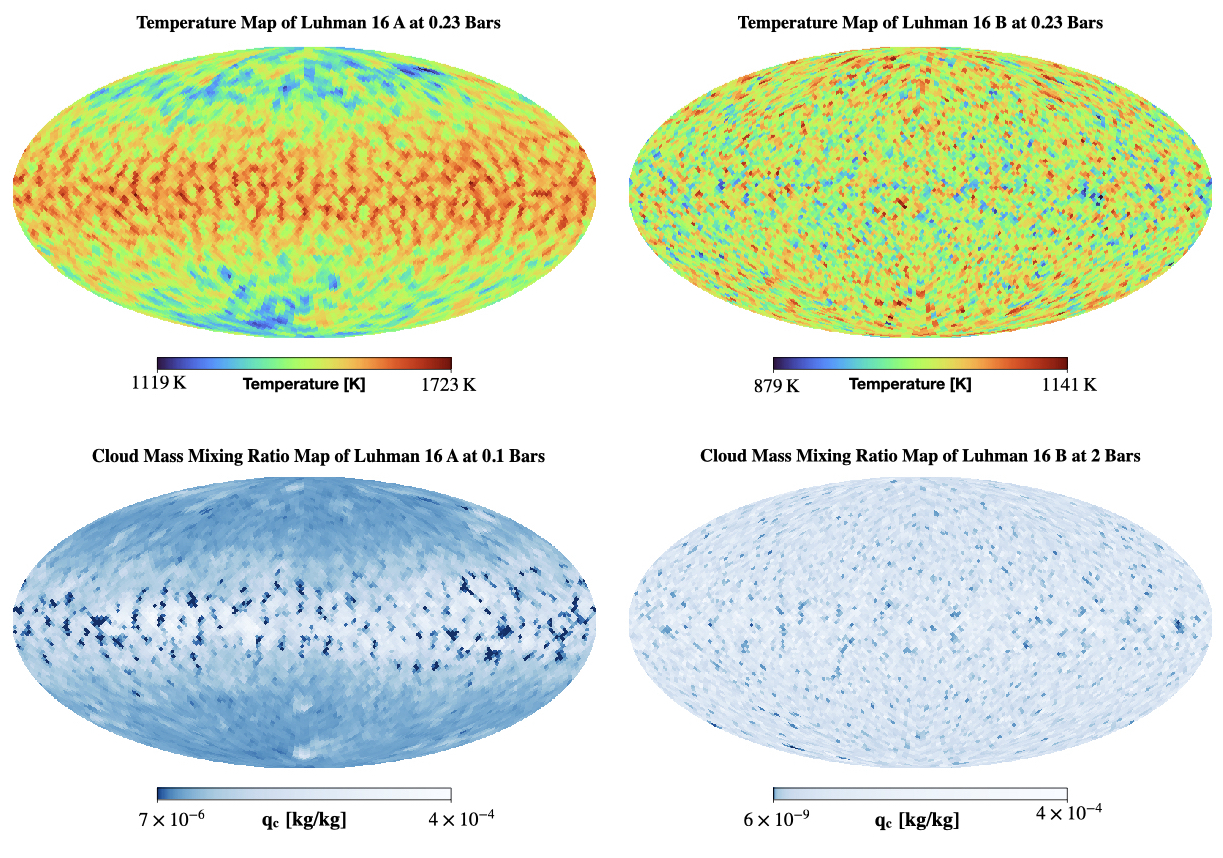}
  
  \caption{The top panel shows the global temperature pattern from the GCM runs at a pressure of 0.23 Bars for Luhman 16 A (top left panel) and Luhman 16 B (top right panel). The bottom panels show the cloud mass mixing ratio ($q_{c}$) at a pressure of 0.1 Bars for Luhman 16 A (bottom left panel) and at a pressure level of 2 Bars for Luhman 16 B (bottom right panel). {\bf Main Point} - The equator is hotter and more cloudy than the poles for Luhman 16 A; furthermore, the size of vortices in Luhman 16 A and B models are different due to these objects' different rotation rates.}
\label{fig:fig3dstruc}
\end{figure*}

The temperature patterns at a pressure level of $\sim$ 0.23 bars arising across the globe for both the brown dwarfs are shown in Figure \ref{fig:fig3dstruc}. This is a representative pressure chosen to be within the enstatite cloud for both objects.  The central band of comparatively hotter regions is very evident for Luhman 16 A whereas for Luhman 16 B such a distinct central hot band is missing at this pressure level. As Luhman 16 A has been assumed to be a slower rotator than Luhman 16 B, horizontal length scales of various vortices and features for Luhman 16 A are larger than Luhman 16 B. The Rossby deformation radius is inversely proportional to the coriolis parameter ($f$) and guides the typical length scales of vortices \citep{tanshowman20} and as a result this difference in length scales of vortices arises in our models.  

The longitudinally averaged temperature-pressure (\emph{T-P}) profile for both Luhman 16 A and B is shown in Figure \ref{fig:fig1dstruc} with colored solid lines starting from the equatorial regions to the polar regions. The \emph{T-P} profiles in the deep interior ($\ge 10$ bars) all converge to the same adiabat due to the onset of convection at such high pressures in both the objects. For Luhman 16 A, the equatorial regions are hotter than the polar regions at lower pressures ($\le 1$ bars). This behaviour can be attributed to cloud radiative feedback. Equatorial regions for both the objects are relatively more cloudy than the polar regions as can be seen in Figure \ref{fig:fig3dstruc} left lower panel. Larger cloud opacities scatter and absorb more thermal flux and as a result heat up the atmospheric columns more in the equatorial regions than the polar regions.

The thermal structure of Luhman 16 B is more homogeneous compared to Luhman 16 A as can be seen in both Figure \ref{fig:fig3dstruc} and \ref{fig:fig1dstruc}. This is related to the cloud optical depth variation for both the objects across the globe. Cloud particle sizes show higher variations across the globe in Luhman 16 A than Luhman 16 B which can be seen both in Figure \ref{fig:fig3dstruc} lower panels and Figure \ref{fig:fig1dstruc} (dotted lines). This causes the cloud radiative feedback on Luhman 16 B to be more homogeneous than Luhman 16 A leading to a compartively homogeneous thermal structure in Luhman 16 B than Luhman 16 A.

\subsubsection{The 3D Cloud Structure}\label{sec:modelcld}
The GCMs treat the condensation and particle formation of MgSiO$_3$ (enstatite) for both the objects. MgSiO$_3$ is a dominant cloud species for objects with effective temperatures for Luhman 16 A and B and has been chosen here as a representative silicate. However, other species like Fe clouds might also be important sources of scattering in the atmosphere but this is not expected to critically effect the cloud circulation patterns as long as the cloud opacities are greater than the gas opacities. We discuss this further in \S \ref{sec:disc}. 

\begin{figure*}
  \centering
  \includegraphics[width=1\textwidth]{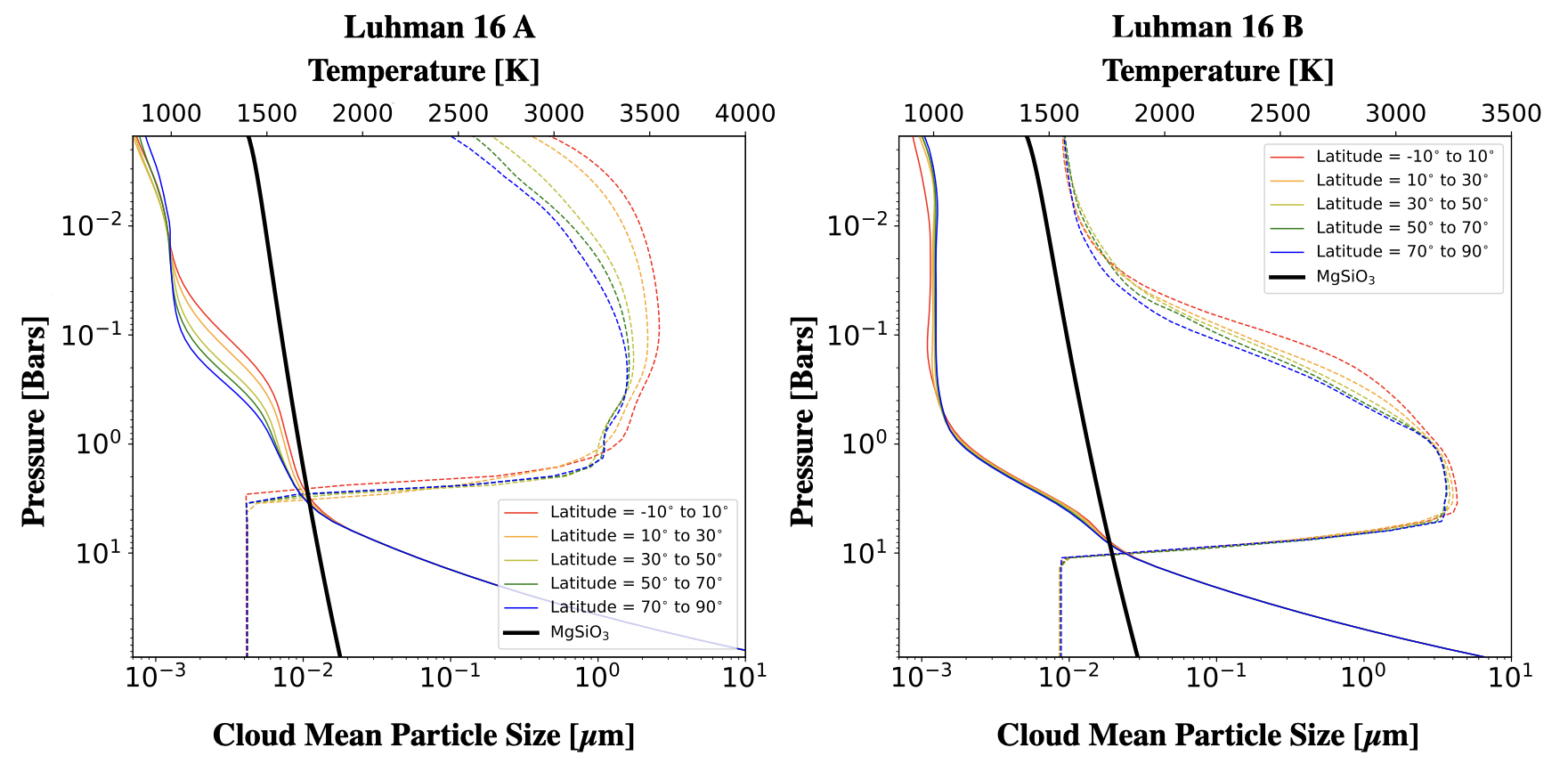}

  \caption{The left panel shows the longitudinally averaged Temperature-Pressure (\emph{T-P}) profiles for Luhman 16 A with solid lines. Various colored lines depict the average \emph{T-P} profile of latitudinal bands each with a angular width of 20$^{\circ}$ starting from the equator (red) to the pole (blue). The black solid line shows the phase-boundary curve for enstatite from \citet{visscher10} assuming solar abundance. The dotted colored lines depict the longitudinally averaged cloud particle sizes in  \,$\mu$m from the equator to the poles. The right panel depicts the same for Luhman 16 B. }
\label{fig:fig1dstruc}
\end{figure*}

The condensing species in the GCM -- enstatite, does not exist in the vapor phase. But it condenses through a chemical pathway involving a reaction between gaseous magnesium, water vapor and SiO \citep{visscher10}. This reaction can only occur across a pressure-temperature curve which we refer to as the phase-boundary curve for enstatite. This phase-boundary curve for enstatite has been treated similarly to a condensation curve for species which condense directly from their vapor phase (like H$_2$O) in our GCM. The gaseous vapor phase Mg in our GCM condenses to enstatite cloud particles when the local mass mixing ratio of the condensible vapor is larger than the local saturation vapor mixing ratio ($q_s$). On the other hand, when the local cloud mixing ratio ($q_c$) exceeds the saturation vapor mixing ratio ($q_s$), it evaporates and turns back to vapor. The local saturation vapor mixing ratio $q_s$ is determined by the temperature-pressure-dependent phase boundary  curve of enstatite given by \citet{visscher10} using solar abundance and has been shown in Figure \ref{fig:fig1dstruc} with black solid lines. The detailed implementation for condensates is the same as that in \citet{tanandshowman19}.


Figure \ref{fig:fig3dstruc} bottom left panel shows the $q_c$ at 0.1 Bar pressure for our models of Luhman 16 A and the bottom right panel shows the $q_c$ at 2 Bar pressure for our models of Luhman 16 B. The pressure levels are chosen here for highlighting the equator to pole trends of the cloud structure in the GCMs near the base of the cloud decks for each object. The pressure dependence of the clouds and their equator to pole trends can be seen in Figure \ref{fig:fig1dstruc}. The sizes of cloud vortices decrease from the equator towards the pole for both the objects due to the variation of the coriolis parameter $f$ with $\phi$ across the globe. Also, Luhman 16 A has larger cloud patterns than Luhman 16 B due to the difference in their rotation periods, as has been discussed earlier in the context of temperature patterns. 

We determine the mean cloud particle size of each layer using \citep{tanshowman20},
\begin{equation}\label{eq:particlesize}
    r_c=\left( \dfrac{3q_c}{4{\pi}N_c{\rho_c}}\right)^{1/3}
\end{equation}
where $q_c$ is the cloud mass mixing ratio and $\rho_c$ is the material density of our cloud species which has been specified in Table \ref{table:planetparams}.

The longitudinally averaged cloud particle sizes are shown for different latitudes from the equator to the pole for Luhman 16 A (left) and Luhman 16 B (right) in Figure \ref{fig:fig1dstruc} with dotted lines. As has been noted in \citet{tanshowman21}, faster rotation leads to vertically thinner cloud decks for Luhman 16 B compared to Luhman 16 A. The typical cloud particle sizes for the Luhman 16 A model is $\sim$ 1  \,$\mu$m whereras that for Luhman 16 B is $\sim$ 5  \,$\mu$m. This is analogous to a low value of $f_{sed}$ -- vertically thicker clouds with smaller particles (Luhman 16 A model) and higher value of $f_{sed}$ -- vertically thinner clouds with larger particles (Luhman 16 B model) -- within the framework of the \citet{ackerman2001cloud} cloud model.  \citet{Buenzli15} found that spectroscopic variability data for the A component is best fit with a f$_{\rm sed}$ = 2 model whereas the B component data could be best fit with a superposition of a hotter model with thinner clouds of f$_{\rm sed}$ = 3 and a colder model with thicker clouds of f$_{\rm sed}$ = 1. In addition, our GCM finds that the cloud thickness decreases towards the pole from the equator due to variation of the coriolis parameter $f$ with latitude. Table \ref{table:planetparams} shows that the number of cloud particles per dry air mass ($N_c$) which has been assumed to be larger for Luhman 16 A than Luhman 16 B as the former is classified to be a L dwarf and the later a T dwarf \citep{kniazev2013,burgasser13}. This results in larger typical cloud particle sizes for Luhman 16 B than the Luhman 16 A model.

\subsection{Radiative Transfer with \texttt{ARTES}}

We use the publicly available 3D Monte Carlo radiative transfer code \texttt{ARTES}\footnote{https://github.com/tomasstolker/ARTES}\citep{stolker17} for post-processing the 3D atmospheric structure calculated from the GCMs. The radiative transfer formalism of \texttt{ARTES} involves emission of photon packages from each grid cell of the three dimensional atmosphere grid. These photons are stochastically  absorbed or scattered as they travel through the atmosphere ultimately hitting the detector once they escape. The initial emission direction of the photons are sampled from a probability distribution function which is biased towards emitting photons in the radially outward direction. The probability distribution function \citep{stolker17} p($\zeta$) is,

\begin{equation}\label{eq:emissionpdf}
    p(\zeta)=\dfrac{\sqrt{1-{\epsilon}^2}}{\pi(1+{\epsilon} \cos \zeta)}
\end{equation}

where $\zeta$ is the emission direction and varies between 0 (radially downward) and $\pi$ radians (radially outwards) and $\epsilon$ is the asymmetry parameter for the emission which varies between 0 and 1. A higher asymmetry parameter ($\epsilon$) leads to a larger number of photons emitted initially towards the radially outwards direction. Since we are mainly concerned about the outgoing thermal emission from the atmosphere, we use $\epsilon$= 0.8 for all our radiative transfer calculations. This is a valid approximation because scattering events which can change the direction of photons emitted to the radially downward directions to radially upward directions are possible but unlikely.

Each photon has a stokes vector of the form,
\begin{equation}\label{eq:stokesvec}
    S = 
    \begin{pmatrix}
    I \\
    Q \\
    U \\
    V 
    \end{pmatrix}
\end{equation}

where ${\pi}I$ is the thermal flux. The degree of linear polarization ($P$) and angle of the polarization vector ($\chi$) for each photon is given by, 

\begin{equation}\label{eq:pol}
    P = \dfrac{\sqrt{Q^2 + U^2}}{I}
\end{equation}

\begin{equation}\label{eq:polangle}
    \chi = \dfrac{1}{2} \arctan \left(\dfrac{U}{Q}\right)
\end{equation}

Each scattering event leads to the rotation of the stokes vector for each photon using a matrix multiplication operation involving the scattering matrix and a rotation matrix depending on the scattering direction of the photon. This operation is described in detail in \citet{stolker17}. The final stokes parameters of the escaped photons are then projected on the detector. The additive nature of the stokes vectors allows the projected stokes vectors to be simply added to calculate the disk-integrated stokes vector of the object. This disk-integrated stokes vector is then used to calculate the disk-integrated flux, degree of linear polarization and angle of linear polarization using Equation \ref{eq:pol} \& \ref{eq:polangle}.

We use 1$\times$10$^{11}$ photons for all of our calculations to maintain sufficient accuracy ($\le$ 10 $\%$ Monte Carlo error) in the calculated thermal and polarized flux. For computational purposes, we regrid the GCM models into a 60$\times$85$\times$171 grid cells ($r\times\theta\times\phi$) for both Luhman 16 A and B. We assume a detector with 100$\times$100 pixels for our calculations. We ensure that our resolution is high enough to preserve the overall circulation patterns of the original GCM runs described in \S \ref{sec:model_gcm} for both bodies.

\subsection{Opacities and Scattering Matrices }

We use the molecular and continuum opacity database available in open-source code \texttt{PICASO} \citep{batalha19,batalha_zenodo_opacities} for our calculations.
Molecular opacities from H$_2$O \citep{barber06high,tennyson2018exomol}, CH$_4$ \citep{yurchenko_2014,yurchenko13vibrational}, NH$_3$ \citep{yurchenko11vibrationally}, CO \citep{li15rovibrational}, PH$_3$ \citep{sousa14exomol}, H$_2$S \citep{azzam16exomol}, CO$_2$ \citep{HUANG2014reliable}, Na \& K \citep{Ryabchikova2015}, TiO \citep{Schwenke98} and VO  \citep{McKemmish16} are included. Collision-induced absorption opacities from H$_2$-H$_2$ \citep{collision-inducedmartin}, H$_2$-He, H$_2$-N$_2$, H$_2$-H, H$_2$-CH$_4$, H-electron bound-free, H-electron free-free and H$_2$-electron interactions are also included. We use the average \emph{T-P} profile of the entire globe to calculate the molecular abundances and atmospheric opacities of these molecules assuming thermochemical equilibrium with solar values for C/O ratio and metallicity. For computational efficiency, we neglect the variation of the equilibrium molecular abundances due to the changing \emph{T-P} profile across radial columns of the atmosphere. As we focus on polarization at a particular wavelength, rather than precision spectroscopy across a broad wavelength range, this simplification is warranted for this particular work. We discuss this further in \S \ref{sec:disc}.

\begin{figure}
  \centering
  \includegraphics[width=0.45\textwidth]{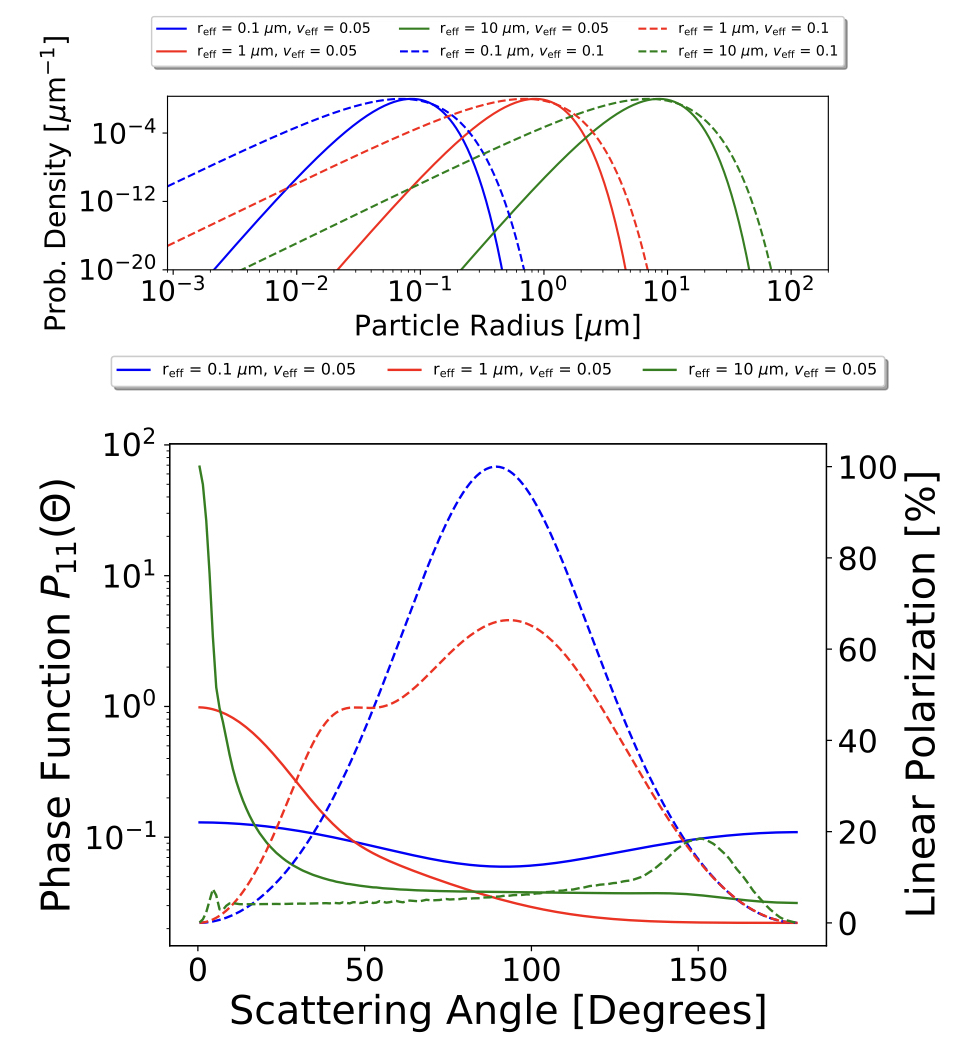}
  \caption{The top panel shows the cloud particle size distributions for different values of effective radius ($r_{\mathrm{eff}}$) and effective variance $v_{\mathrm{eff}}$. The solid lines depict the particle size distributions with $v_{\mathrm{eff}}$= 0.05 and  $r_{\mathrm{eff}}$ of 0.1 \,$\mu$m (blue solid line), 1 \,$\mu$m (red solid line) and 10 \,$\mu$m (green solid line). The dashed lines show wider particle size distributions with $v_{\mathrm{eff}}$= 0.1 and $r_{\mathrm{eff}}$ of 0.1 \,$\mu$m (blue dashed line), 1 \,$\mu$m (red dashed line) and 10 \,$\mu$m (green dashed line). The bottom panel left vertical axis correspond to phase function ($P_{11}(\Theta)$). The solid lines show $P_{11}(\Theta)$  calculated from the particle size distributions shown in the top panel with the same colors. The percentage of linear polarization as a function of the scattering angle is shown with dashed lines for each of the particle distributions (right vertical axis). {\bf Main Point}- Gas and clouds scatter infrared light very differently and single scattering polarizations are very sensitive to particle sizes.}
\label{fig:figphase}
\end{figure}

We assume that the cloud particle sizes follow a gamma distribution \citep{hansen71,stolker17} although the GCM runs includes only single-sized particles. The cloud particle size distribution is given by,
\begin{equation}\label{eq:sizedist}
    n(r)=Cr{^{(1-3v_{\mathrm{eff}})/v_{\mathrm{eff}}}}e^{-r/v_{\mathrm{eff}}r_{\mathrm{eff}}}
\end{equation}
where $r_{\mathrm{eff}}$, $v_{\mathrm{eff}}$ and $C$ are the effective radius, effective variance and the normalization constant, respectively. Figure \ref{fig:figphase} top panels show the particle size distribution for various combinations of $r_{\mathrm{eff}}$ and $v_{\mathrm{eff}}$. The solid lines show the particle size distributions when $v_{\mathrm{eff}}$=0.05 and $r_{\mathrm{eff}}$ is 0.1 (blue), 1 (red) and 10 (green) \,$\mu$m. The dotted lines show the distributions when $v_{\mathrm{eff}}$=0.1 and $r_{\mathrm{eff}}$ is 0.1 (blue), 1 (red) and 10 (green) \,$\mu$m. Other cloud models like \citet{ackerman2001cloud} assume log-normal distribution of particle sizes. The gamma particle size distributions differ from log-normal distribution because of its asymmetry. It represents clouds where probability density is higher for larger particles near $r_{\mathrm{eff}}$ with a broader distribution of smaller particles.

For each $(r,\theta,\phi)$ grid point we assume the r$_{\mathrm{eff}}$ to be the cloud particle radius from Equation \ref{eq:particlesize} and $v_{\mathrm{eff}}$ to be 0.05. We then calculate the particle size distribution in 80 radial bins and calculate the absorption opacity, scattering opacity and the scattering matrix elements in each of these radial bins using Mie particle scattering theory. We obtain the complex refractive indices of amorphous magnesium silicates similar to enstatite in composition from \citet{scott96} and use it for all our cloud opacity and scattering matrix calculations. These opacities and matrix elements are then integrated over the size distribution to obtain the total cloud opacities and scattering matrix elements for that grid point. This calculaion is done using the Mie calculation module in \texttt{ARTES} \citep{stolker17}. The scattering matrices for the gas and cloud particles are calculated using the following form \citep{hansen74,stolker17},

\begin{equation}\label{eq:scatmatrix}
    S = 
    \begin{pmatrix}
    P_{11}(\Theta) & P_{12}(\Theta) & 0 & 0  \\
    P_{12}(\Theta) & P_{22}(\Theta) & 0 & 0 \\
    0 & 0 & P_{33}(\Theta) & P_{34}(\Theta)  \\
    0 & 0 & -P_{34}(\Theta) & P_{44}(\Theta)  
    \end{pmatrix}
\end{equation}

where $\Theta$ is the angle between the directions of the incident and the scattered light. The $P_{11}(\Theta)$ component represents the phase function of the scattering when normalized appropriately such that $\dfrac{1}{2\pi}\int_{0}^{\pi} P_{11}(\Theta)sin(\Theta) \,d\Theta=1$.
As the gas particle sizes are much smaller than infrared wavelengths, this matrix is calculated in the Rayleigh scattering regime for the gas particles where,
\begin{align*}
    P_{11}(\Theta)&= \dfrac{3\pi}{4}(1+cos^2\Theta)  & P_{12}(\Theta)&= \dfrac{3\pi}{4}(1+cos^2\Theta) \\
    P_{22}(\Theta)&=  P_{11}(\Theta)  & P_{33}(\Theta)&= \dfrac{3\pi}{4}(2cos\Theta) \\
    P_{44}(\Theta)&=  P_{33}(\Theta)  & P_{34}(\Theta)&= 0 \\
\end{align*}

This Rayleigh scattering matrix represents a situation where the incident wavefront is scattered symmetrically in the forward and backward directions. Since the cloud particles scatter infrared photons in the Mie particle scattering regime, the cloud scattering matrix is calculated using the full Mie theory with the  Mie calculation module in \texttt{ARTES}. The Mie scattering matrix for micron sized cloud particles produces asymmetrically scattered wavefronts. This has been shown in Figure \ref{fig:figphase} bottom panel. The solid lines show the  Mie particle scattering phase function ($P_{11}(\Theta)$) for a wavelength of 1.6 \,$\mu$m  for the three particle size distributions shown in the top panel with solid lines corresponding to $v_{\mathrm{eff}}$=0.05 and $r_{\mathrm{eff}}$ is 0.1 (blue), 1 (red) and 10 (green) \,$\mu$m. The dotted lines show the polarization percentage from a single scattering event as a function of scattering angle. The phase function in solid blue shows that particles scatter light symmetrically in the forward and the backward direction with less light scattered at an angle of 90$^{\circ}$ when the $r_{\mathrm{eff}}$ is $\sim$ 0.1 \,$\mu$m. The polarization percentage for $\sim$ 0.1 \,$\mu$m particles peak at a scattering angle of 90$^{\circ}$ which is a typical Rayleigh scattering behaviour. Particle size distributions with typically larger particles like $r_{\mathrm{eff}}$ of 1 or 10 \,$\mu$m show much higher amounts of forward scattering compared to backward scattering. With increasing particle sizes, the degree of linear polarization also departs from the Rayleigh regime behaviour.

\section{Results}\label{sec:res}
\subsection{Luhman 16 A}\label{sec:luh16a}
\begin{figure*}
  \centering
  \includegraphics[width=1\textwidth]{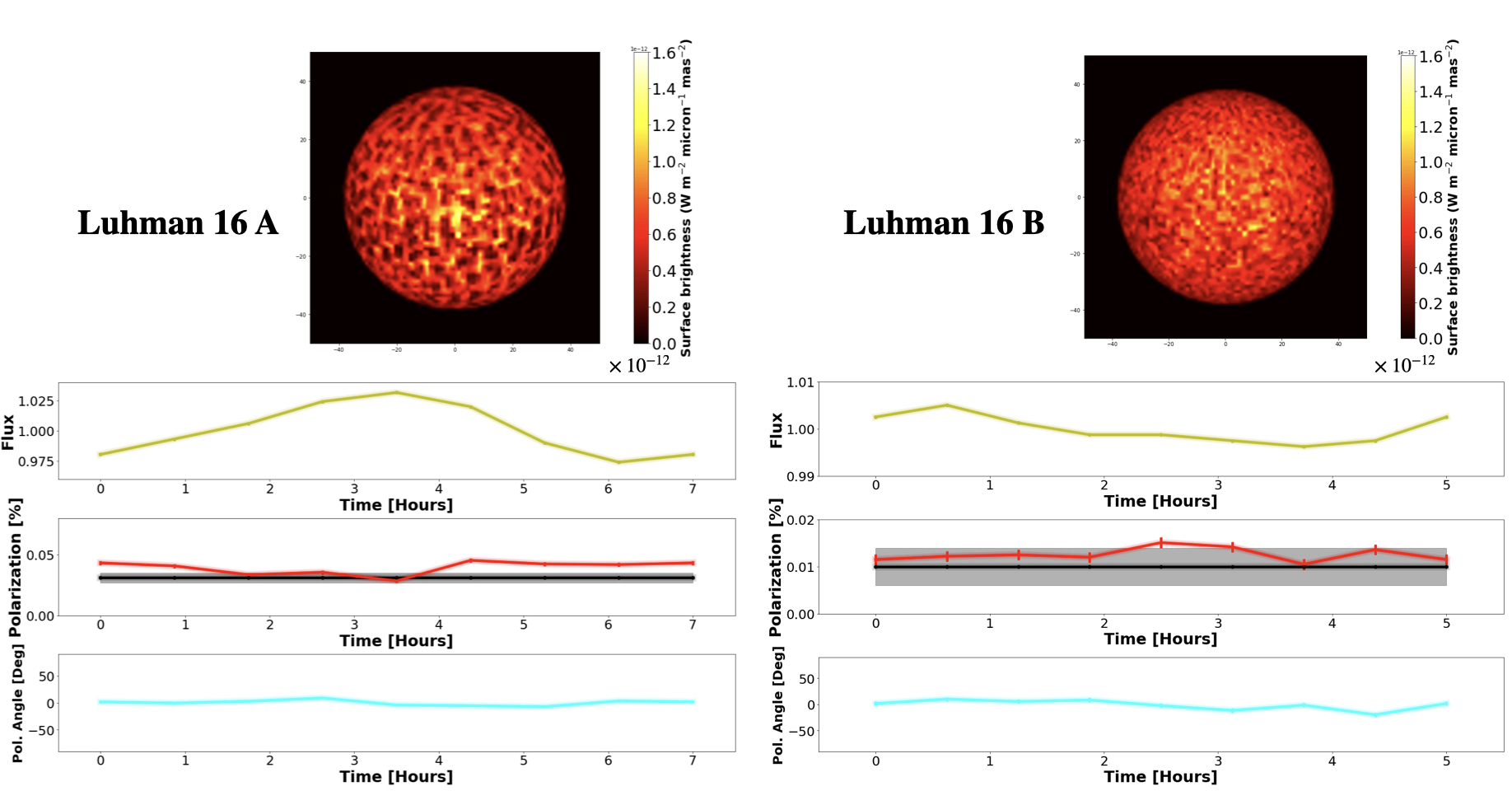}
  \caption{The left panel shows the radiative transfer result for Luhman 16 A. The top left panel shows the surface brightness map for the object. The second panel in the left shows the variability of thermal flux at 1.6  \,$\mu$m with time as the object rotates with a period of 7 hours. The third left panel shows the linear polarization percentage vs. time with the solid red line. The solid black line and the shaded region shows the measured polarization for the object. The bottom left panel shows the variation of the angle of polarization of the object with time using the cyan line. The same plots for Luhman 16 B (5 hour period) has been shown in the right panels. {\bf Main Point} - Our models match the degree of linear polarization well for both the objects but do not reproduce the observed optical variability amplitude of Luhman 16 B.  }
\label{fig:figpolab}
\end{figure*}

We first calculate the thermal flux and linear polarization percentage for Luhman 16 A at 1.6  \,$\mu$m. This choice of wavelength is motivated by the recent measurement of the polarization of Luhman 16 A in the H-band by \citet{mmb20}. The measured polarization is 0.031$\pm$0.004\%. For Luhman 16 A, we assume a rotation period of 7 Hours \citep{apai21} and a rotationally-induced oblateness of 0.003 which is expected from a Luhman 16 A like body with a rotational period of 7-8 hours \citep{mmb20}. We assume an inclination of 28$^{\circ}$ for our initial calculation (0$^{\circ}$ is equator-on hereafter). This is motivated by the recent measurements of inclination to be $\le$ 28$^{\circ}$ for Luhman 16 A by \citet{apai21}. We explore other inclinations closer to the equator-on view as well and describe the results from them. We calculate the polarization and flux signal from our Luhman 16 A GCM every 52.5 minutes as the object rotates for a total of one rotation period. This interval is chosen such that we repeat our calculation after every 45$^{\circ}$ rotation of the object along the spin axis. Figure \ref{fig:figpolab} left panels show the surface brightness map of Luhman 16 A, thermal flux variability, linear polarization \% variability and angle of polarization variability from top to bottom, respectively.

The surface brightness map for Luhman 16 A shows the thermal flux at 1.6  \,$\mu$m from each pixel at a single time snapshot as the object is rotating. The central relatively cloudy band is readily visible to be slightly off-centred in the image due to the assumed 28$^{\circ}$ inclination of the object. The cloudy regions emit less thermal flux and appear darker because they scatter/absorb more light compared to the clear regions. The flux variability of the object for one complete rotational period is shown with the yellow solid line. The peak-to-peak variability of the thermal flux is about 5\% which mainly arises from the varying cloud patterns across the globe in our GCM. \citet{Buenzli15}  measured a 4\%  peak-to-peak variability of Luhman 16 A within the 0.8-1.15 \,$\mu$m wavelength range. Our results do not reproduce the \citet{Buenzli15} measurement within the 1$\sigma$ uncertainty. Our calculated flux variability is also larger than the measurements of \citet{apai21} who infer a typical peak-to-peak variability amplitude of 2.2\% for Luhman 16 A using TESS light curves of the system in the wavelength range of 0.6-1.0 \,$\mu$m. But these differences can be due to difference in wavelength of our calculations (1.6  \,$\mu$m) and the observed wavelengths as the variability amplitude can vary strongly with wavelength \citep{Buenzli15}. \citet{biller13} measured the H-band peak-to-peak amplitude for the unresolved Luhman 16  binary to be $\sim$ 4\%. However, the H-band amplitude of the Luhman 16 A component alone remained below the detection limits of their observations. Differences in variability amplitudes can also arise if our circulation model is not capturing the `true' cloud distribution and/or particle sizes in these objects.     

The time varying modeled polarization signal is shown with the red solid line and is compared to the measured time averaged signal in black for Luhman 16 A \citep{mmb20}. The uncertainty in the measured polarization is shown with the shaded black region. The modeled polarization signal agrees well with the time averaged measurement. But importantly, the degree of polarization also shows significant variation within 0.03\% to 0.045\% within one rotational period.

Variability of linear polarization has been previously observed in cool M--dwarf stars \citep{milespaez15} in optical wavelengths. However, the origin of those polarization signals are still unclear and can be due to the strong magnetic field of the cool dwarf, presence of a dusty disk around the star or presence of non-uniformly distributed dust grains in the stellar photosphere. \citet{dekok11} also found time variable polarization signal from a brown dwarf model with a dusty hot spot in its atmosphere. We also simulate the polarization signals for more equator-on inclinations $\sim$20$^{\circ}$ and $\sim$0$^{\circ}$ (equator-on). The modeled time-averaged signal is $\sim$0.04\% and $\sim$0.07\% respectively for the two inclinations. Both these signals are much higher than the measured value. This indicates that inclination of Luhman 16 A is very close to 28$^{\circ}$ if our GCM represents its atmosphere correctly.  We discuss this further in \S \ref{sec:disc}.

The disk-integrated polarization signal is also anti-correlated to the disk-integrated flux from the object. This is because scattering from cloud particles are primarily responsible for the polarized flux in infrared wavelengths, and hence a higher disk-integrated flux hints towards lower cloud coverage at that time instant and can give rise to lower disk-integrated signal.

The angle of linear polarization also varies significantly for Luhman 16 A as shown in the bottom left panel of Figure \ref{fig:figpolab}. The angle of polarization varies between -7$^{\circ}$ to 8$^{\circ}$ within one full rotational period. This kind of rotation of the plane of the polarization vector was also seen in models of brown dwarfs with a single dusty hot spot in \citet{dekok11}. Our simulations suggest that the angle of polarization must show significant time variations if they arise from such circulation patterns. These polarization observations integrate over time scales of several hours which are comparable to the rotational period of these objects. We sum up the disk-integrated stokes parameters from each of our time steps to calculate the disk-integrated degree of linear polarization and angle of the polarization vector when time-integrated over one rotational period. We find that the time integrated degree of linear polarization is 0.038 \% and the angle of the linear polarization is -0.28$^{\circ}$. As has been outlined in \citet{mmb20} and \citet{stolker17}, the polarization vector aligns itself with the spin axis if the disk-integrated polarization is dominated by equatorial bands of clouds and if oblateness dominates the polarization signal then the polarization vector is perpendicular to the spin axis. For Luhman 16 A, the polarization vector in all time steps as well as the disk-integrated polarization is aligned very close to the spin axis which suggests that our modeled polarization is mainly dominated by the equatorial bands of clouds in the GCM. The measured angle of linear polarization for Luhman 16 A by \citet{mmb20} is -32$^{\circ}$$\pm4$$^{\circ}$ relative to the North direction on the sky. This angle is not comparable to our model results since the on-sky orientation of the spin-axis is unknown.

We check the sensitivity of the disk-integrated polarized signal to the typical cloud particle sizes of our GCM models. For Luhman 16 A, the typical cloud particle sizes are $\sim$ 1  \,$\mu$m (we assume a gamma distribution of cloud particles) but we also simulate instances when the typical cloud particle sizes are $\sim$ 10  \,$\mu$m and $\sim$ 0.5  \,$\mu$m. Although changing the cloud particle sizes should also alter the circulation pattern in principle due to the dependence of cloud opacities on particle sizes, we neglect that subtlety here. Figure \ref{fig:particlesize} shows the dependence of the disk-integrated polarization signal for both of our Luhman 16 A (top panels) and Luhman 16 B (bottom panels) models. The top left column shows the surface brightness map of the Luhman 16 A model when the typical cloud particle sizes are $\sim$ 1  \,$\mu$m and the polarization percentage matches the observed polarization well. But when the particle sizes are increased to $\sim$ 10  \,$\mu$m the surface brightness increases but the polarization signal declines by about $\sim$ 10 times. The polarization signal increases slightly if the cloud particle sizes are taken to be about $\sim$ 0.5  \,$\mu$m. This indicates that the typical cloud particle sizes for Luhman 16 A, if the circulation patterns are representative of the reality, lies between $\sim$ 0.5-1  \,$\mu$m.

\subsection{Luhman 16 B}\label{sec:luh16b}

The simulated surface brightness map of Luhman 16 B from the GCM models described in \S \ref{sec:model_gcm} is shown in Figure \ref{fig:figpolab} top right panel. We have assumed an inclination of 26$^{\circ}$ for Luhman 16 B motivated by the results presented in \citet{mmb20}. The time variability of thermal flux, degree of linear polarization and angle of linear polarization are shown in the right three panels from top to bottom, respectively.  The modeled peak-to-peak variability for Luhman 16 B is about $\sim$ 1\% which is much less than the observed Luhman 16 B typical H-band variability amplitude of about 13\% measured by \citet{biller13}. This shows that our GCM model is inadequate for explaining the variability of Luhman 16 B. This might happen due to the presence of a large cloud/cloudless feature on the globe of Luhman 16 B which is not present in our GCM run. 


The variability of the disk-integrated polarization signal for the Luhman 16 B model matches well with the measured polarization of 0.010\%$\pm$0.004\% \citep{mmb20} for an assumed inclination of 26$^{\circ}$ for the object. We explore the equator-on configuration for the Luhman 16 B model and find a disk-integrated polarization percentage of $\sim$0.03\% which is $\sim$ 3 times larger than the measured value. So, our model prefers an inclination of 26$^{\circ}$ for Luhman 16 B in contrast to a much more equator-on configuration suggested by the variability analysis in \citet{apai21}. However, we take this opportunity to reiterate that the detection significance of the measurement is only 2.5-sigma.

\begin{figure*}
  \centering
  \includegraphics[width=1\textwidth]{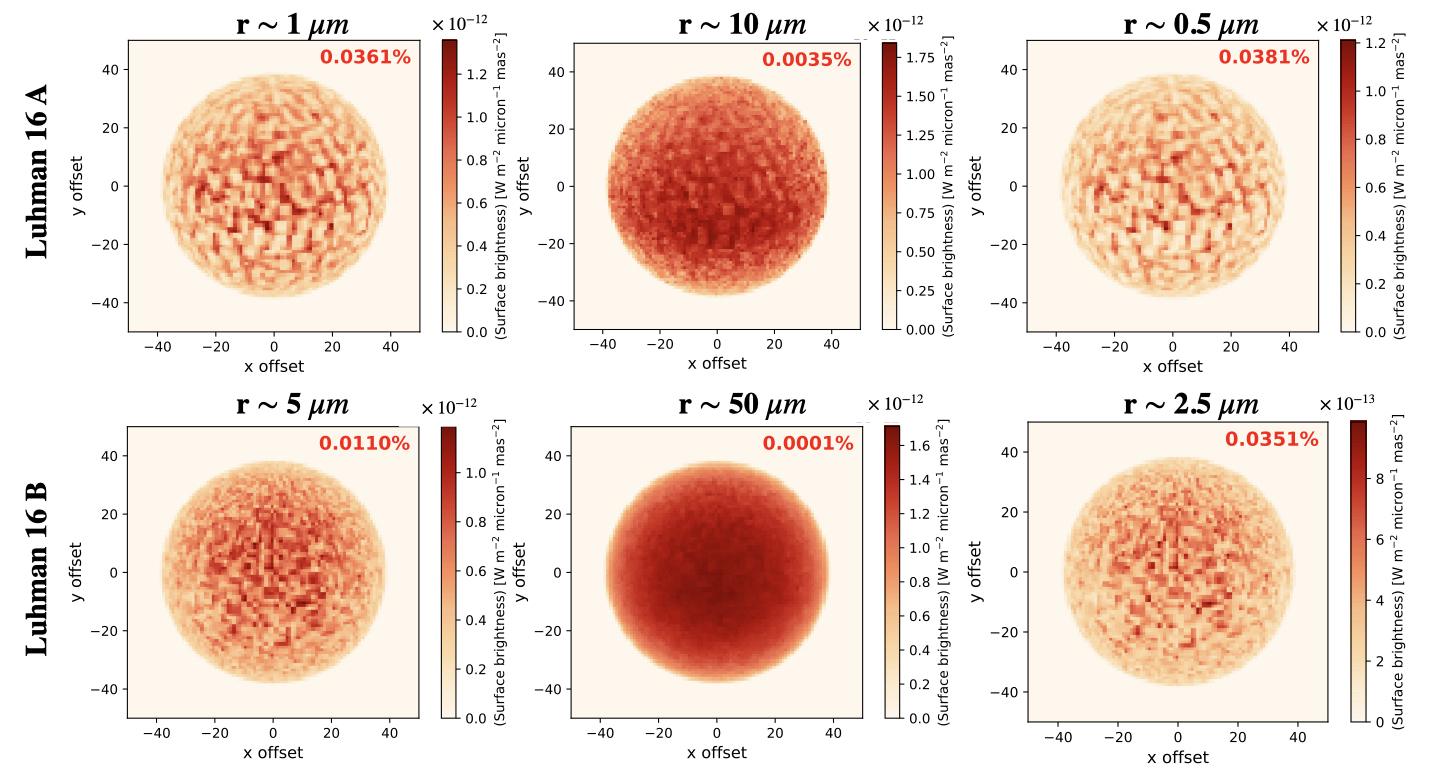}
  \caption{The top left, middle and right panel shows surface brightness map for Luhman 16 A if the typical cloud particle sizes were 1 \,$\mu$m (original GCM), 10 \,$\mu$m and 0.5 \,$\mu$m. The disk-integrated polarization for each case has been shown in the top right of each plot window. The bottom left, middle and right panel shows surface brightness map for Luhman 16 B if the typical cloud particle sizes were 5 \,$\mu$m (original GCM), 50 \,$\mu$m and 2.5 \,$\mu$m.  Please note that the color scheme is inverted. {\bf Main Point} - Luhman 16 A should have typical cloud particle sizes betwen 0.5-1 \,$\mu$m whereas Luhman 16 B should have cloud particle sizes of about 5 \,$\mu$m.}
\label{fig:particlesize}
\end{figure*}

The typical cloud particle sizes for Luhman 16 B GCM is $\sim$5 \,$\mu$m. The disk-integrated polarization signal at 1.6 \,$\mu$m for cloud particle sizes of $\sim$2.5  \,$\mu$m and $\sim$50  \,$\mu$m are found to be 0.0351\% and 0.0001\%, respectively. The calculated signal with particles of $\sim$50 \,$\mu$m sizes is  $\sim$10 times lower than the measured signal whereas the signal with typically $\sim$2.5 \,$\mu$m clouds particles is $\sim$3 times higher than the measurement. This has been shown in Figure \ref{fig:particlesize} lower panels. This suggests that the typical cloud particles sizes for Luhman 16 B is $\sim$5 \,$\mu$m. 

 The angle of linear polarization also varies significantly for Luhman 16 B between 10$^{\circ}$ and -20$^{\circ}$. This suggests that our polarization calculations for Luhman 16 B are dominated by the equatorial band of clouds in our Luhman 16 B circulation model. We find the time-averaged degree of linear polarization from our Luhman 16 B model is 0.011 \% by integrating the calculated signal over one rotation period of the object. This is in good agreement with the measured H--band polarization of 0.01 $\pm$ 0.004 \%  for Luhman 16 B. This time-integrated polarization vector projects an angle of -1.56$^{\circ}$ from the spin-axis of the object. The measured time averaged angle of the polarization vector is  73$^{\circ}$ \citep{mmb20} relative to the North direction in the sky and hence is not directly comparable to our modeled angles since the on-sky orientation of the spin-axis is unknown.


\subsection{Modeling of Polarization Signal with Band Models}

Single and multiple band models have often been used to model polarization signal from objects with non-uniformity in cloud coverage \citep{mmb20,stolker17}. These have been motivated by observed band-like cloud patterns in Jupiter. Here, we have tried to explore if such models are adequate for calculations of disk-integrated polarization signals from objects like those modeled by our GCM runs or if we need circulation models like the one used in this work for interpreting future polarization observations from non-uniformly cloudy objects. We first assume that the real nature of circulation on a brown dwarf/exoplanet is represented by our circulation models. We then create models with multiple bands of clouds (``band models") and test if they can reproduce the same polarization signal as we calculate from our GCM runs for Luhman 16 A and B. We assume that the cloud structure of our band models for Luhman 16 A and B is the same longitudinally averaged cloud structure shown in Figure \ref{fig:fig1dstruc}. From the equator to the pole, each of the cloud profiles shown in Figure \ref{fig:fig1dstruc} left panel for Luhman 16 A forms a band with a latitude span of 20$^{\circ}$ in our band model for Luhman 16 A. Similarly, we construct our band model for Luhman 16 B with the averaged cloud profiles shown in Figure \ref{fig:fig1dstruc} right panel forming bands with latitude spans of 20$^{\circ}$. The gas opacities used for the radiative transfer for the GCM are also used for the band models of the corresponding objects to ensure similarities in all conditions between the GCM models and the band models except for the detailed circulation patterns dominating the smaller length scales than the global jets. We compute the polarization signal from the band models for Luhman 16 A and Luhman 16 B using the same oblateness and inclinations as has been used for the computations reported in \S \ref{sec:luh16a} and \ref{sec:luh16b}.

\begin{figure*}
  \centering
  \includegraphics[width=1\textwidth]{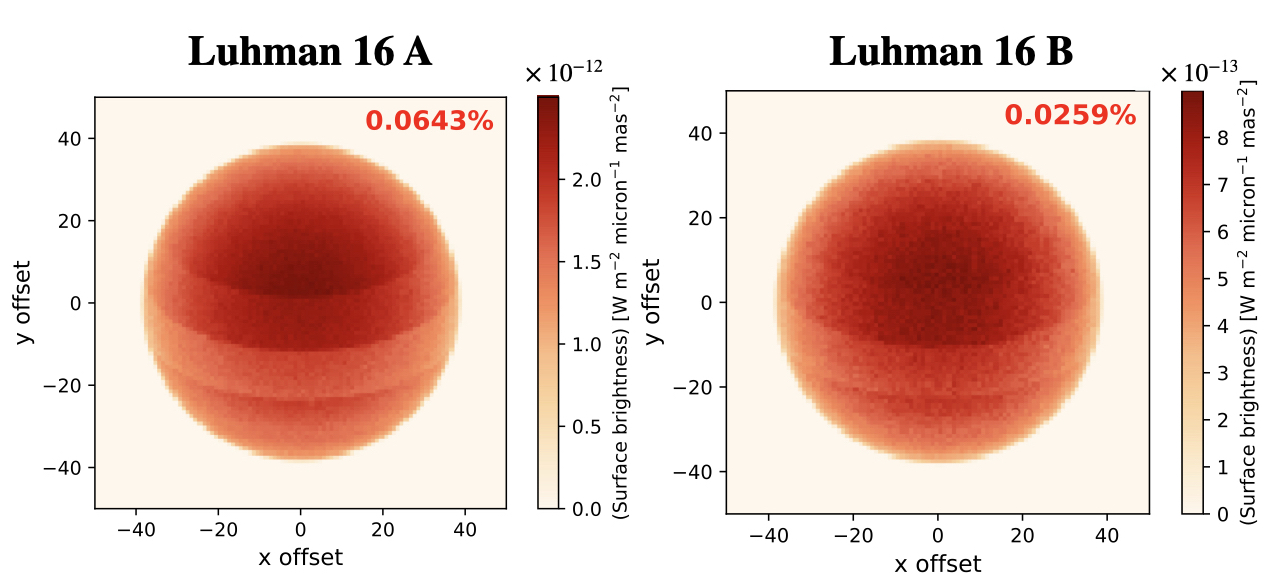}
  \caption{Left panel shows the surface brightness map of the simple band model for Luhman 16 A along with the 1.6  \,$\mu$m disk-integrated polarization percentage denoted at the top in red. Right panel shows the same for the band model of Luhman 16 B. Please note that the color scheme is inverted. {\bf Main Point}- Band models tend to overestimate polarization signal. }
\label{fig:mock}
\end{figure*}

Figure \ref{fig:mock} shows the surface brightness map of the multiple band model for Luhman 16 A and the right panel shows the same for the  multiple band model of Luhman 16 B. The disk-integrated polarization signal from this band model for Luhman 16 A is found to be 0.0643\%. This is almost twice the signal which was calculated from the GCM for Luhman 16 A where the polarization was found to be $\sim$0.038\%. For Luhman 16 B, the calculated polarization from the band model is 0.0259\% in contrast to the circulation model polarization percentage of $\sim$0.011\%. The linear polarization vectors from the band models of both the objects are aligned along the spin-axis as has also been found from the GCM models. So in both the cases the band models overestimate the disk-integrated polarization compared with the GCM models even if other properties like inclination, oblateness, cloud optical properties, gas opacities and the thermal structure are almost alike between the band and the circulation models. This hints towards the significant influence of smaller scale vortices across the globe of these objects, which are present only in the circulation models, on the disk-integrated polarizations. Even if both Luhman 16 A and B circulation models have a central equatorial band of clouds, a band model with single or multiple cloud bands might still overestimate the polarization signal from the object. These smaller-scale vortices of cloudy and cloud free regions tend to decrease the overall disk-integrated polarization signal.

\section{Discussion}\label{sec:disc}
We first discuss some of the caveats, implications and future follow up of our work and then summarize our findings in \S \ref{sec:conc}.

\subsection{GCM Parameter Choices}
Our circulation models use the computational framework described in \citet{tanshowman20}, which necessarily involves some simplifications. For example, the GCM models and our subsequent radiative transfer calculations only consider the condensation of one cloud species (enstatite) which is an important cloud species in the temperature regime of our interest. Other cloud species like Fe and  Olivine are also suggested to be important sources of cloud opacity, and can alter the cloud radiative feedback, or cloud optical properties, for the object. The choice of the condensation species itself is not very critical as long as the cloud optical depths exceed the gaseous optical depths \citep{tanshowman21}. However, different cloud species like Fe might follow a different particle size distribution which can have an effect on the polarization signal calculations.

For our models presented here,  we use a particular time snapshot of the circulation structure of these objects after they reach a quasi-steady state. However, these circulation patterns may continue to evolve further in time and therefore investigation of the long-term photometric and polarization variability may prove fruitful as well.

The circulation models used here can be sensitive to other assumptions as well, such as the drag strength deep in the brown dwarf atmosphere. The GCMs presented here have been calculated in the ``strong drag" regime but this drag strength is highly uncertain as the interaction between the deeper interior and the atmosphere is not properly characterized to date \citep{tanshowman21}. This work uses these GCM runs for post-processing them with the comprehensive radiative transfer tool \texttt{ARTES} which allows us to make direct comparisons with observations but there remains ample scope for further exploration of the parameter space of the circulation models themselves. In future work, we will explore a wider range of GCM parameter space for Luhman 16 A and B, and generic brown dwarfs, to examine the sensitivity of our radiative transfer results to the assumptions outlined above.  The combination of GCM outputs and post-processing for the calculation of polarization and spectra appears to be a powerful tool.

\subsection{Spherical Cloud Particles}

We assume cloud particles to be spherical throughout this study for simplicity, as it is typical in the field, and use Mie particle scattering theory for calculation of the cloud scattering properties. However, non-spherical particles can have significantly different scattering phase functions especially at back-scattering angles as has been investigated in detail by \citet{mi07000g}.  However, they show that the difference in the cloud optical properties like optical depth, single scattering albedo and asymmetry parameter between a non-spherical and projected-area equivalent spherical distribution of particles is at maximum $\sim$ 10\%. This suggests that considering non-spherical particles in our GCM will not significantly alter the cloud radiative forcing compared to the spherical particles assumed in this work.

Since the phase functions at back-scattering angles differ significantly between spherical and non-spherical distributions of particles \citep{mi07000g}, they can affect the disk-integrated polarization signals calculated in this work. This has also been shown in \citet{stolker17} where the departure from spherical particles has been treated with the distribution of hollow spheres approach \citep{min05}. We use a spherical cloud particle shape assumption to avoid additional complexity at this time, in our already complex modeling.  We defer treating non-spherical cloud particles for future work as more polarimetric observations are obtained in near future.

Lastly, optical constants for mineral aerosol particles can themselves be temperature dependent \citep{Zeidler_2015}, which we ignore here.  Neither the infrared emission spectra or the polarimetric measurements of brown dwarfs are constrained enough at this point of time to investigate these effects.

\subsection{Improvements to Radiative Transfer and Opacities}

All our calculations are done at a single wavelength of 1.6 \,$\mu$m whereas the polarimetric measurements with which we compare our models are often done in a broader photometric wavelength bands (H--band for Luhman 16 A and Luhman 16 B). In this work, we do not perform our radiative transfer calculations for multiple wavelengths within the photometric wavelength band of observations for making our comparison with data in order to avoid the much larger computational time required. But integrating the disk-integrated polarization across a photometric band might change the predicted polarization signal from GCMs especially if there are sharp molecular absorption or emission windows within the photometric band in question.

In future modeling work over a broader wavelength range, to fit both emission spectroscopy as well as polarization, additional computational efforts should be put into opacity variations around the globe.  Gaseous opacities depend on both the local temperature and pressure, and recall here  that we use the gaseous opacities for the globally averaged \emph{T-P} profile for our radiative transfer.  We make this approximation because our main focus in this work is the calculation of the polarization signal and gas molecules are not important scattering sources in infrared wavelengths. The only way they effect the resultant infrared thermal polarized flux is with their absorption opacity. As the maximum fluctuation in temperature across the globe at a certain pressure level for Luhman 16 A is about $\sim$200 K and that for Luhman 16 B is $\sim$20 K, the temperature dependance of gas opacity across the globe will be a small effect and will not affect the net-disk integrated polarization signal significantly, but may become important in future work.

\section{Conclusions}\label{sec:conc}
We have coupled pioneering 3D GCM calculations for brown dwarfs Luhman 16 A and B with the vector Monte Carlo radiative transfer code \texttt{ARTES}, in order to calculate the polarization signal in three dimensions.  We summarize our findings in the form of answers to the specific questions we promised to answer/explore in \S \ref{sec:intro} here.
\begin{enumerate}
    \item We find that General Circulation Models, with nearly global enstatite clouds, can be used to explain the observed H--band disk-integrated degree of linear polarization signals from cloudy brown dwarfs Luhman 16 A and Luhman 16 B. We calculate the degree of linear polarization for Luhman 16 A to be 0.038 \% and the measured value is 0.031\% $\pm$ 0.004\%. For Luhman 16 B, our calculated degree of polarization is 0.011\% and the measured polarization is 0.010\% $\pm$ 0.004\%.  
    
    \item The typical variability amplitude observed in Luhman 16 A is also reproduced from the thermal flux calculated from our GCM for Luhman 16 A. However, our Luhman 16 B GCM underestimates the variability amplitudes from the observations of Luhman 16 B.  However, in all cases the observed variability for Luhman 16 A was not at 1.6 $\mu$m, the wavelength we model here.

    \item If our circulation patterns are representative of the circulation patterns on these objects, then we find that the preferred spin-axis inclination for Luhman 16 A is $\sim$ 28$^{\circ}$ whereas that for Luhman 16 B is $\sim$ 26$^{\circ}$. We find that the disk-integrated polarization signals calculated from GCMs also decline as one moves from equator-on to pole-on configurations.
    
    \item We also find that decreasing cloud particle sizes generally increases the disk-integrated polarization signal. For our GCM to match the observations for Luhman 16 A, we find that the typical cloud particle sizes should be within 0.5--1  \,$\mu$m. Whereas, for Luhman 16 B, cloud particles should be around 5  \,$\mu$m as both 2.5  \,$\mu$m or 50  \,$\mu$m overestimate and underestimate the disk-integrated polarization, respectively.
    
    \item Simpler cloud band models with one or multiple bands have often been used to interpret polarization measurements from non-uniformly cloud covered objects. We find that even if we use cloud profiles very similar to the GCMs keeping every other important parameter such as oblateness and inclination the same, these band models tend to overestimate the disk-integrated polarization signal compared with the GCMs. This suggests that smaller scale vortices tend to decrease the disk-integrated polarizations in brown dwarfs and are important factors influencing the polarized emission. This also indicates that if the real cloud circulation of any object is like the GCMs then band models might not be adequate to interpret disk-integrated polarization measurements from them.

\end{enumerate}

 In our follow-up work we will expand to multi-wavelength polarization predictions, both for Luhman 16 A and B, and for brown dwarfs generally over a larger parameter phase space, given the larger data sets being accumulated \citep{jensenclem20,holstein21}. We will also explore a wider range of GCMs for Luhman 16 A and Luhman 16 B, to assess whether the observed broadband photometric variability in the visible and near infrared, as well as the polarization signal, can be interpreted in the framework of one model. Also, our cloudy GCMs for Luhman 16 A and B do not reproduce the multiple band like circulation patterns inferred by \citep{apai21} for Luhman 16 B from TESS observations. Further exploration of our GCM parameter space for these objects and investigating the circulation patterns and the disk-integrated polarization signals arising from them will significantly progress our understanding of brown dwarf clouds.

\section{Acknowledgments}
SM thanks the UC Regents Fellowship award for supporting him for this work. JJF acknowledges support of the Simons Foundation. RJC acknowledges support from NSF AAG grant ATI2103241.  We acknowledge use of the lux supercomputer at UC Santa Cruz, funded by NSF MRI grant AST 1828315. We thank Maxwell Millar-Blanchaer and the anonymous referee for valuable comments that helped to improve the paper. We also thank Tomas Stolker for help and guidance about the \texttt{ARTES} code and Xi Zhang for useful discussions.

{\it Software:} ARTES \citep{stolker17}, PICASO \citep{batalha19}, pandas \citep{mckinney2010data}, NumPy \citep{walt2011numpy}, IPython \citep{perez2007ipython}, Jupyter \citep{kluyver2016jupyter}, matplotlib \citep{Hunter:2007}

\bibliography{sample63}{}
\bibliographystyle{apj}



\end{document}